# Impact of intrinsic localized modes of atomic motion on materials properties

M. E. Manley

*Lawrence Livermore National Laboratory, Livermore, California 94551*

**Abstract**

Recent neutron and x-ray scattering measurements show intrinsic localized modes (ILMs) in metallic uranium and ionic sodium iodide. Here, the role ILMs play in the behavior of these materials is examined. With the thermal activation of ILMs, thermal expansion is enhanced, made more anisotropic, and, at a microscopic level, becomes inhomogeneous. Interstitial diffusion, ionic conductivity, the annealing rate of radiation damage, and void growth are all influenced by ILMs. The lattice thermal conductivity is suppressed above the ILM activation temperature while no impact is observed in the electrical conductivity. This complement of transport properties suggests that ILMs could improve thermoelectric performance. Ramifications also include thermal ratcheting, a transition from brittle to ductile fracture, and possibly a phase transformation in uranium.

**Keywords**: Intrinsic localized modes, point defects, diffusion, mechanical properties, thermal conductivity.

1. **Introduction**

More than two decades ago it was proposed that nonlinearity and discreteness could cause some vibrational energy to spontaneously localize in small regions (just a few atoms across) of a perfect crystal [1, 2]. Since that time theoretical work has shown that these intrinsic localized modes (ILMs) – also called discrete breathers or lattice solitons – should be ubiquitous in nature, occurring in any number of lattice dimensions and in a variety of physical systems [3, 4]. Indeed, experiments have demonstrated ILMs in a range of driven systems, including two-dimensional Josephson-junction arrays [5, 6], optical wave-guide arrays [7], microwave driven antiferromagnets [8], and acoustically driven micromechanical resonator arrays [9]. Only recently, however, has evidence emerged of ILMs forming in conventional three-dimensional crystals in thermal equilibrium; first in elemental α-U metal above ~450 K [10, 11] and then in a simple ionic crystal, NaI, above ~555 K [12]. In thermal equilibrium ILMs are expected to form randomly stabilized by configurational entropy, much like vacancies only with lower activation energy [2]. The activation energy originates with large amplitude local fluctuations needed for the ILM dynamics to escape the plane wave frequency spectrum, plus the cost of any local structural distortions [11]. To understand the formation mechanism of an ILM, note that in an anharmonic lattice the frequency of a local fluctuation depends on amplitude and, for sufficiently large amplitudes, this frequency may fall outside the plane wave spectrum, either above the phonon cutoff or within a spectral phonon gap. Once formed these off-resonance dynamic hotspots do not couple well to any of the plane wave modes, allowing the energy to remain confined in a local mode [3]. In the present article, the ramifications of these equilibrium ILMs are examined for materials properties. The cases of α-U and NaI, where the ILM activation temperatures are known [10-12], are used to examine general theoretical predictions and ideas from the literature on the role ILMs play in thermal expansion, diffusion, radiation damage, ionic conductivity, mechanical deformation, thermal transport, and possibly a phase transformation. Practical ramifications and potential future applications of these ILM manifestations are also discussed.

## 2. Thermal Expansion

A link between ILMs and thermal expansion was hinted at in the pioneering 1988 paper of Sievers and Takeno when they noted that each ILM should be accompanied by a structural relaxation (dc distortion) [2]. It follows that the thermal activation of increasing numbers of ILMs with increasing temperature should have the aggregate effect of changing the crystal size, thereby contributing to thermal expansion.

A more detailed assessment of the dc component of an ILM was determined in 1997 by Kiselev and Sievers [13] using molecular dynamic simulations on NaI [13]. In these simulations they found that for sufficiently large initial amplitudes ILMs polarized along the [111] crystallographic directions were stable [13]. The ac-eigenvector of the stable ILM is reproduced in Fig 1. The dominant component of the ILM is the motion of the light central Na atom, with smaller movements extending out into neighboring unit cells. The existence, size (~1.2 nm), optic-mode-like structure factor, and [111] polarization of the ILM were verified in 2009 using inelastic neutron and x-ray scattering [12]. The dc-eigenvector of the ILM, also shown in Fig. 1, has a distinct form. Along the ILM polarization direction there is a local contraction in the lattice while in the orthogonal directions the lattice expands. The cubic symmetry of NaI prohibits observation of this detail since equivalent crystallographic directions average out to preserve symmetry. The expected net effect, however, is an enhancement of the average thermal expansion since the expansions exceed contractions. The thermal expansion of NaI, measured along the [100] axis of a single crystal by Rapp and Merchant [14], is shown in Fig. 2. Consistent with prediction, the displacement-temperature curve shows a small slope increase near the temperature where the ILMs in NaI are first observed to form (~555 K [12]), as shown in Fig. 2.

The orthorhombic symmetry of $\alpha$-U provides for a more detailed look at the impact of ILM symmetry on thermal expansion since the ILMs have atomic displacements polarized along the unique [010] direction [10]. The anisotropic thermal expansion of $\alpha$-U was measured along all three crystallographic directions on a single crystal by Lloyd [15]. Figure 3a shows the results along with an indication of the ILM activation temperature, ~450 K [10]. There is a significant change in the thermal expansion behavior starting around the ILM activation temperature. Along the ILM polarization direction, [010], the thermal expansion changes from positive to negative, while the two orthogonal directions show enhanced expansions. This is precisely the pattern expected for the thermal activation of ILMs with dc-eigenvectors similar to those predicted for NaI, Fig. 1, as noted in Ref. [16]. This enhanced anisotropy is more convincing evidence of ILM involvement in thermal expansion; especially the development of negative thermal expansion along the ILM polarization axis since this is difficult to explain by conventional anharmonic mechanisms.

Perhaps the most surprising evidence of ILM involvement in the thermal expansion of uranium is obtained from a comparison of the thermal expansion determined from bulk crystal dimensions with that determined from lattice spacing deduced by x-ray diffraction. The effect of ILM activation on the volume expansion of a uranium crystal, Fig. 3b, is consistent with NaI in that there is an increase in the crystal volume expansion rate above the ILM activation temperature; compare solid black lines in Fig. 2 and Fig. 3b. By contrast, however, the thermal expansion of uranium determined using x-ray diffraction by Bridge [17] shows the opposite effect. Rather than an increase there is a slight decrease in the volume expansion rate with ILM activation, blue line in Fig. 3b. This decrease in the x-ray determined volume expansion rate near the ILM activation temperature was first identified by W. D. Wilkinson [18] in 1958; the effect is more pronounced in this data reproduced in the bottom panel of Fig. 3b. The occurrence of more expansion in the bulk crystal than predicted from the lattice spacing determined by x-ray diffraction is indicative of a local contribution. The positions of the Bragg peaks are determined by long-range periodic features, while local distortions only serve reduce the intensities of the Bragg peaks [19]. This does not, however, explain the decrease in the expansion rate of the x-ray determined volume. This peculiarity may actually be a manifestation of the energy concentrating property of ILMs [3, 4]. Scattering measurements on both $\alpha$-U and NaI indicate that associated with ILM formation are intensity losses and decreases in the rates of softening (decreasing frequency with increasing temperature) in associated normal phonon modes [10, 12]. These changes are consistent with some vibrational energy being drawn away from these normal modes. Since these modes reside in regions away from ILMs, and since these modes contribute to thermal expansion, a loss of energy in them should result in a decrease in the thermal expansion rate of material away from the ILMs, i.e. material probed by the x-ray diffraction measurements. Hence, although ILMs increase the overall rate of thermal expansion they may actually decrease the rate slightly in regions of the crystal away from the ILMs. This interpretation is illustrated in Fig. 4, where the ILMs appear as dynamic hotspots (red) with associated local distortions, while the surrounding lattice appears effectively cooler (yellow) and less expanded.

Interestingly, the large anisotropy in the high-temperature thermal expansion of α-U, which appears to be associated with ILM activation, has a dramatic side effect. Polycrystalline ingots textured along the [010] direction (ILM polarization direction) exhibit a dramatic swelling effect under thermal cycling [18]. As described in Ref. [18], after cycling 300 times from 323 K to 823 K an ingot grew along the [010] textured axis from 2 inches to 11.5 inches, developing a twisted-rope-like appearance. This thermal ratcheting effect does not occur in single crystals [18] and is thought to result from microstrains that develop owing to the large anisotropy in the thermal expansion of individual crystallites, which here we associate with the activation of ILMs. When discovered this effect, along with a radiation swelling effect, resulted in the abandonment of α-U metal by the nuclear industry [20]. As Lander *et al.* put it in a 1994 review, "*This surely represents one of the most rapid changes of technology driven by basic research*" [20]. According to Cahn in *The Coming of Materials Science* [21], as of 2000 just a few thermal reactors fueled by metallic uranium remain in use in Britain.

3. **Kinetic Processes: Diffusion, Radiation Damage, and Ionic Conductivity**

In 1995 it was suggested by Russell and Collins that non-equilibrium ILMs are generated during irradiation and carry energy over large distances away from the damage cascade [22]. Subsequent theoretical work lent some support to this idea by showing that it was possible for ILMs to travel over macroscopic distances at specific velocities along certain crystallographic directions [23, 24]. A possible connection of this idea to diffusion came with the 1D simulations of Cuevas et al. [25] showing that collisions between propagating ILMs and vacancies can cause the vacancies to hop between sites. Another development was the experimental work of Abrasonis, Möller, and Ma in 2006, who reported that the diffusion of interstitial N in stainless steel is enhanced under $Ar^+$ irradiation at distances ~1 micron beyond the ion penetration depth [26]. After ruling out other causes, they argued that the effect originates with energy transmitted by long-distance traveling ILMs generated in the irradiated region. These arguments suggest three ILM properties: (1) a long mean free path for traveling ILMs, (2) the possibility of generating non-equilibrium ILMs by irradiation, and (3) a mechanism for ILMs to enhance or influence a diffusion process. The latter two are considered in this section.

For both α-U and NaI non-equilibrium ILMs were generated experimentally using both inelastic neutron and x-rays scattering [11, 12]. In the process it was shown that these non-equilibrium ILMs are essentially equivalent to those found in thermal equilibrium [11, 12]. Specifically, the results show that the ILMs are generated by large amplitude fluctuations that mirror the modes themselves in both polarization and reciprocal space structure, and that the required creation energy is consistent with the thermal activation energy of the equilibrium ILMs [11]. Hence, irradiation induced ILM generation is supported by measurement. Furthermore, since non-equilibrium ILMs appear to be similar to the thermally activated ILMs, any influence on diffusion should extend to thermally activated ILMs.

Striking evidence of an ILM effect on diffusion can be found in a large unexplained decrease in the thermal activation energy for the fast-diffusion of interstitial $Cu^+$ in NaI [27, 28]. The decrease in activation energy occurs at the ILM activation temperature and was observed independently by two groups, one using optical studies of diffusion profiles [27] and the other using Nuclear Magnetic Resonance (NMR) measurements of the jump frequencies [28]. A log plot of the NMR jump frequency data versus inverse temperature is shown in Fig. 5, along with an indication of the ILM activation temperature for NaI [12]. The activation energy, given by the negative slope, abruptly decreases from about 1.2 eV to 0.6 eV [28] above the ILM activation temperature. Recent 1-D simulations indicate that ILM-interstitial collisions can cause interstitial atoms to hop between sites [29], but this does not explain these observations since such collision events would only add to the jump frequency above the ILM temperature, not diminish it. A better insight comes from studies of ILMs in micromechanical cantilever arrays. When a short "impurity" cantilever is introduced to an array ILMs can become attracted to and trapped at the "impurity" cantilever [9]. An ILM bound to a diffusing impurity in a crystal could physically alter the local environment, thereby altering the diffusion mechanism rather than adding to it.

Also shown in Fig. 5 are the activation plots for Na self diffusion, near bottom, and vacancy diffusion, near top. Neither of these rates appears to be affected by ILM activation. Theoretical work by Cuevas et al. [25] indicates that ILM-vacancy collisions in 1-D models can cause vacancies to hop between sites but only if a threshold energy is exceeded. Another property that depends on the migration of vacancies, however, is the ionic conductivity. As shown in Fig. 6a, ionic conductivity data on pure NaI taken from Ref. [30] exhibits a sharp transition at the ILM activation temperature. The fact that there is a transition is not unusual; all of the alkali halides show a similar transition from low to high temperature behavior [30]. What is unusual, however, is how low this transition temperature is relative to melting in NaI, about 0.59 of its melting temperature. For comparison LiCl, which has the same crystal structure and a similar atomic mass ratio (about 5), undergoes this transition at 0.94 of its melting temperature [30]. So perhaps the activation of ILMs reduces this transition temperature. Returning to the ILM-vacancy collision simulations of Cuevas et al. [25], one way this might occur is through ILM assisted vacancy migration under the force of an applied electric field. Even though ILM-vacancy collision processes do not assist free vacancy diffusion (Fig. 5), the force of the applied electric field might lower the energy requirement enough for such collisions to move vacancies during the ionic conduction process.

Another observation is the change in the activation energy for the ionic conductivity of NaI doped with Cd as reported in Ref. [31] and reproduced in Fig. 6b. Like with Cu diffusion (Fig. 5) the activation energy for the ionic conductivity of Cd impurities in NaI decreases sharply just above the ILM activation temperature (Fig. 6b). Since the movements of impurities during diffusion and ionic conductivity are related, a similar mechanism may be occurring for both, possibly the trapping of ILMs at the impurity sites as discussed above for Cu diffusion. The more important point here, however, is that again a change is occurring at the activation ILM temperature. Combined the three sets of observations tend to rule out some accidental coincidences with the ILM temperature. The first set of data, Fig. 5, tends to rule out a coincidental vacancy process energy and Fig. 5 and Fig. 6b combined tends to rule out a coincidental energy barrier for a specific impurity (Cu and Cd should be different).

The effect of ILM activation on diffusion in uranium is more difficult to investigate because the ILM activation temperature is too low for much diffusion to occur (the homologous temperature, $T/T_{melting}$, at ILM activation is 0.32 in α-U). However, there are indications that kinetic processes activate coincidentally with the activation of the ILMs. First, carbon distributions dissolved within the grains of uranium metal change with time above the ILM temperature but not below [32], showing that the mobility of carbon interstitial atoms initiates at the same temperature ILMs activate. Second, radiation damage effects due only to displaced uranium atoms anneal out above the ILM activation temperature [18], indicating that dislodged uranium atoms relax back into their equilibrium positions at the same temperature ILMs activate. Combined, these two very different observations, point to an enhancement of kinetic processes in uranium with the activation of ILMs (a third piece of evidence on uranium will be provided in the next section on mechanical deformation). These effects might be understood qualitatively in terms of the Cuevas ILM-defect collision models [25, 29], where ILM collisions knock displaced uranium atoms back into equilibrium positions and start to knock around impurities.

In summary, interstitial diffusion processes ($Cu^+$ in NaI and C in α-U) appear to be influenced by ILM activation, while some evidence of an influence on self-atom, substitutional impurity (Cd), and vacancy migration is found but only in the externally perturbed states; α-U under irradiation and NaI under an applied electric field during ionic conduction. These effects are not fully understood but probably involve both ILM-defect collision processes [25, 29] and the trapping of ILMs by impurities [9].

4.  **Mechanical Deformation and Fracture**

In the discovery paper on ILMs in α-U it was noted that a loss of mechanical ductility occurs at the ILM activation temperature [10]. It was suggested that the ILMs may be interacting with the defects involved in the deformation process [10]. The data referenced is reproduced in Fig. 7 and shows the percent elongation at failure as a function of temperature as reported by Taplin and Martin [33]. Although there is a change from increasing elongation with temperature to a plateau above the ILM temperature, percent elongation data is difficult to interpret by itself because there are many stages to the failure process. Here these stages are dissected further.

A typical stress-strain curve leading to the failure of a tensile test of α-U is shown at the top of Fig. 8. The first stage is the elastic limit. A careful inspection of all the elastic constants measured on α-U by Fisher [34] shows no obvious anomaly at the ILM temperature, Fig. 8a. This indicates that there is nothing in long-range elastic binding energy of uranium that is changing at this temperature, consistent with a local effect. The next stage is the yield stress, defined at 0.2% offset from linear behavior. As shown in Figure 8b, the yield stress reported by Holden [18] also shows no anomaly at the ILM activation temperature. So it does not appear to be related to the onset of dislocation glide or twin boundary motion, both of which occur in uranium [35]. Finally, at the end stage, fracture mechanisms do show a significant change at the ILM activation temperature. As indicated in the fracture mechanism map developed by Collins and Taplin [36] and summarized in Fig. 8c, there is a transition from brittle to ductile fracture mechanisms at this temperature. Most interesting, however, is that in the high temperature range all the fractures are thought to be initiated by the formation of very small cavities that grow by diffusion [36]. So it appears that the effect of ILMs on mechanical deformation, rather than operating through the motion of dislocations or twins, is caused by the impact of ILMs on diffusion (see previous section). Hence, a practical ramification of ILM-enhanced diffusion appears to be the transition from brittle to ductile fracture in uranium precipitated by incipient void growth.

## 5. Thermal Conductivity

Just as dislocations represent increments of plastic strain and vacancies represent increments of free volume, ILMs represent increments of kinetic energy or heat. Thermal conductivity therefore appears to be an obvious property to consider. Indeed, the problem of transport by traveling ILMs or lattice solitons has been considered theoretically and shown to have a positive effect in certain 1D model systems [37, 38]. Recent theoretical work has also shown that ILMs can wander through the lattice via a classical process analogous to quantum tunneling, but can also become self trapped [39]. Collisions between two propagating ILMs result in energy being transferred between them, with the more localized excitation taking energy from the less localized excitation [3, 40-45]. This property tends to reinforce the most localized ILMs as they move through the lattice [46]. However, the lack of dispersion (group velocity) in the mode along with its associated local distortion [13] suggests that ILMs move relatively slowly and can act as scattering centers for normal phonons, thereby decreasing thermal conductivity.

A single curve representing the best engineering average of the thermal conductivity of uranium [18] is reproduced in the top panel of Figure 10. There is a marked inflection at the ILM activation temperature. This inflection was noted in the 1950's and associated with the changes in the thermal expansion behavior (bottom panel in Fig. 3b) although at the time both were simply attributed to a possible "*change in the nature of the binding*" [18]. The electrical resistivity, bottom panel in Fig. 9, shows no corresponding change at the ILM activation temperature. The lack of correspondence between the electrical and thermal transport indicates that the inflection is not due to a change in the resistance to the flow of free electron carriers. This electronic contribution is calculated by applying the Wiedemann-Franz law [47] to the electrical resistivity data, where the electronic thermal conductivity is given by $k_{el} = LT/\rho$, where $L$ is the Lorenz number ($2.45 \times 10^{-8}$ W-Ω/K$^2$), $T$ is temperature, and $\rho$ is the electrical resistivity. The electronic contribution, appearing as a dotted line in the top panel of Fig. 9, fits well to the thermal conductivity data above the ILM activation temperature. Below the ILM activation temperature, however, there is a significant additional contribution from the lattice vibrations. At these temperatures the lattice contribution is probably enhanced by a steeply dispersing soft-mode anomaly [48] associated with a charge density wave that forms below 42 K [49]. The underlying electron-phonon coupling may also be contributing to the nonlinearity responsible for the ILMs [10, 50]. The interesting observation here is that this lattice contribution abruptly vanishes above the ILM activation temperature. It is not clear, however, if the suppression of the lattice contribution is caused by phonon scattering from ILMs, by the self trapping of vibrational energy in forming ILMs, or through some indirect process connected to associated changes in the normal modes [10].

Unfortunately, thermal conductivity measurements have not been reported for NaI in the critical region close to the ILM activation temperature. Measurements were made from 80 K to 400 K (around Debye temperature) by Devyatkova and Smirnov [51] and from 760 K to 930 K (near the melting temperature) by Khokhlov, Kodintseva, and Filatov [52]. These results are summarized in Fig. 10. In the lower temperature data there are two regions; below ~120 K corresponds to having only acoustic modes occupied and above ~120 K corresponds to having both acoustic and optic modes occupied [51]. An interpolation of the power law dependence across the ILM temperature overshoots the highest temperature thermal conductivity by about factor of 1.5, suggesting another mechanism may come into play. Power law deviations do occur in other alkali halide salts [53] and perhaps ILMs are involved, but more measurements are needed to establish a connection.

A possible application of the ILM effect on transport, as observed in uranium, might be the development of better thermoelectric materials. Thermoelectric materials are important for power-generating devices designed to convert waste heat into electrical energy and have become the subject of intense interest in recent years [54-59], owing to a renewed interest in energy efficiency. A material's figure-of-merit used to judge its suitability is: $ZT=\alpha^2 \sigma T/k$, where $\alpha$ is the Seeback coefficient, $\sigma$ is the electrical conductivity, and $k$ is the thermal conductivity. Performance is improved by increasing electrical conductivity while minimizing thermal conductivity. Hence, the suppression of the lattice thermal conductivity in the absence of any change in the electrical conductivity with ILM activation (Fig. 10), suggests that ILMs could improve the thermoelectric figure of merit.

## 6. Phase Transformation

In 2004 Bussmann-Holder and Bishop [60] proposed that in a wide class of ferroelectric materials, where coupled electron-ion polarizability provides a strong source of nonlinearity, ILMs may play a role in the development of nanoscale heterogeneity during the ferroelectric transformation. In 2005 they worked with Egami and obtained some experimental evidence in support of this idea [61]. For NaI there is no solid-state phase transformation so its ILMs are clearly not associated with a phase change. For α-U, however, there are several phase transformations. In particular, on heating there is a narrow intermediate tetragonal β-phase followed by a broad phase field of body centered cubic (γ–phase), from which uranium melts [20]. It was noted that the symmetry properties of the ILMs in α-U put the local structure on a path towards the high temperature γ-phase and that this suggests that the ILM may be a precursor to the phase transformation [62]. Recall from section 2 that as ILMs become thermally activated (above ~ 450 K [10]), the lattice parameter of α-U along the polarization direction of the ILM, the b-axis [10], begins to contract while the two orthogonal directions show an enhanced expansion, Fig. 3a. As illustrated in Fig. 11a, these local distortions help drive the symmetry towards hexagonal [62]. Furthermore, as noted in Ref. [62] some of the phonon anomalies associated with the ILM appear to take on hexagonal symmetry as the temperature is increased towards the phase transition. In relation to the high temperature bcc phase this has significance since there is a crystallographic path from α-phase to the high temperature bcc structure that passes through an intermediate hexagonal structure. This path is illustrated in Fig. 11b.

On the one hand, this is perhaps not that surprising since the nonlinearity responsible for the ILMs should be related to real structural instabilities. In this sense it simply explains the ILM orientation. If, on the other hand, the ILMs are more fundamental to the phase transformation mechanism, as argued by Bussmann-Holder and Bishop for ferroelectrics [60, 61], it could be of considerable importance. It reverses perceptions familiar from the soft-mode mechanism [63], where a soft mode develops in a high temperature phase as it is cooled and ultimately lowers the symmetry and internal energy during the phase transformation. The ILMs develop in the low temperature phase, are stabilized by configurational entropy with increasing temperature [2, 10], and act as a precursor to the higher entropy phase [62].

## 7. Conclusions

Knowledge of thermally activated ILMs in $\alpha$-U and NaI sheds new light on the high temperature behavior of materials. Thermal expansion is enhanced, made more anisotropic, and becomes inhomogeneous at a microscopic level. Interstitial diffusion, ionic conductivity, the annealing rate of radiation damage, and void growth all exhibit changes coincident with the thermal activation of ILMs. The influence on these kinetically controlled processes likely involves both ILM-defect collision effects [25, 29] and ILM trapping at impurities sites [9]. These observations combined with previous nonequilibrium ILM creation experiments [11, 12] lend some support to the idea that radiation generated ILMs could enhance or at least influence kinetic processes [22, 26]. The mechanical deformability of uranium is diminished with the appearance of ILMs, but this is not caused by the interaction of ILMs with dislocations or twins. Rather, the loss of deformability appears to originate with fractures nucleated at small voids that begin to form by diffusion above the ILM temperature. This is consistent with ILMs, which are point-like defects in the distribution of kinetic energy, and are expected to have a larger impact on the point defects involved in diffusion than on line (dislocation) or plane (twin) defects. The lattice component of the thermal conductivity of uranium is strongly suppressed at the ILM activation temperature while no impact is observed in the transport of free electron carriers. These transport effects suggest that ILMs might improve the performance of thermoelectric materials. Other practical ramifications of these ILM manifestations include a thermal ratcheting effect that helped redirect a nuclear industry and a transition from brittle to ductile fracture in uranium. Finally, the local symmetry of the ILMs in uranium is also consistent with them being a precursor to a solid-state phase transformation. Apparently, hidden within the atomic vibrations of some materials exist dynamic nonlinear hotspots as important to properties as the well studied defects of the crystal structure.

**Acknowledgements**

Work was performed under the auspices of the U.S. Department of Energy by Lawrence Livermore National Laboratory under Contract DE-AC52-07NA27344. The author would like to thank A. J. Sievers, C. Marianetti, K. Moore, B. Fultz, M. Asta, and J. Cuevas for useful comments.

**Figure Captions**

Figure 1. Eigenvectors of the oscillatory motion, "ac", and static distortion, "dc", of an ILM simulated using model potentials for NaI, after Ref. [13]. The small atoms are Na and the large atoms are I. The ILM oscillations, centered at Na, are polarized along the [111] direction. The local dc distortions include a contraction along the direction that the mode displaces atoms in oscillatory motion and an expansion in the orthogonal direction.

Figure 2. Thermal expansion of NaI as measured from the dilation of a single crystal along the [100] direction, after Ref. [14].

Figure 3. Thermal expansion of α-U. (a) Percent length change along the three principle axes of a single crystal measured by Lloyd, Ref. [15]. (b) Percent change in volume determined from the single crystal data of Lloyd [15], solid black line in top panel, and determined from x-ray diffraction determined lattice spacing by Bridge et al. [17], blue line top panel, and Wilkinson [18], blue line bottom panel. The ILM activation temperature is indicated with the vertical red line.

Figure 4. Simplified illustration of how ILMs may manifest as an inhomogeneous distribution of both lattice vibrational energy and anharmonic thermal expansion. The focusing of the lattice dynamical energy is accompanied by a concomitant focusing of the thermal expansion effect in localized regions.

Figure 5. Activation energy plot for jump frequencies extracted from NMR data on NaI doped with small amounts of Cu+, after Ref [28]. The ILM temperature is indicated with the vertical red line.

Figure 6. Activation energy plot determined from the ionic conductivity of NaI: (a) Pure NaI data taken from Ref. [30], (b) Cadmium doped NaI data with a Cd mass fraction (m.f.) of $33 \times 10^{-6}$, after Ref. [31]. The ILM temperature is indicated with the vertical red line.

Figure 7. Percent elongation at failure of $\alpha$-U pulled in tension, after Ref. [32]. The ILM temperature is indicated with the vertical red line.

Figure 8. Mechanical properties of $\alpha$-U near the ILM temperature. A typical stress-strain curve is shown at the top with relevant properties presented below; (a) elastic constants, after Ref. [33], (b) yield stress, after Ref. [18], and (c) fracture mechanism map, after Ref. [35]. The ILM temperature is indicated with the vertical red line.

Figure 9. Transport properties of $\alpha$-U taken from Ref. [18]. Top panel: Solid line is the measured thermal conductivity; Bottom panel: Measured electrical resistance. The dotted line in top panel labeled $TL/\rho$ is the electronic contribution to the thermal conductivity estimated from the electrical resistivity data in the bottom panel, see text.

Figure 10. Thermal conductivity measured on NaI above and below the ILM activation temperature (indicated by vertical red line). The open circles in the lower temperature range were taken from Ref. [50] and the solid line at the highest temperatures was taken from Ref. [51].

Figure 11. The symmetry of ILMs in α-U. (a) The Brillioun zone viewed down the c-axis as it is distorted by the ILM distortions deduced from the thermal expansion shown in Fig. 6 (Because this is reciprocal space the distortions look inverse, the contraction along [010] appears as an expansion of the zone, etc.). (b) A view of the α-U structure, where the open circles are the top plane and the solid are below, as it is transformed to the bcc phase after passing through an hcp-like hexagonal structure.

NaI

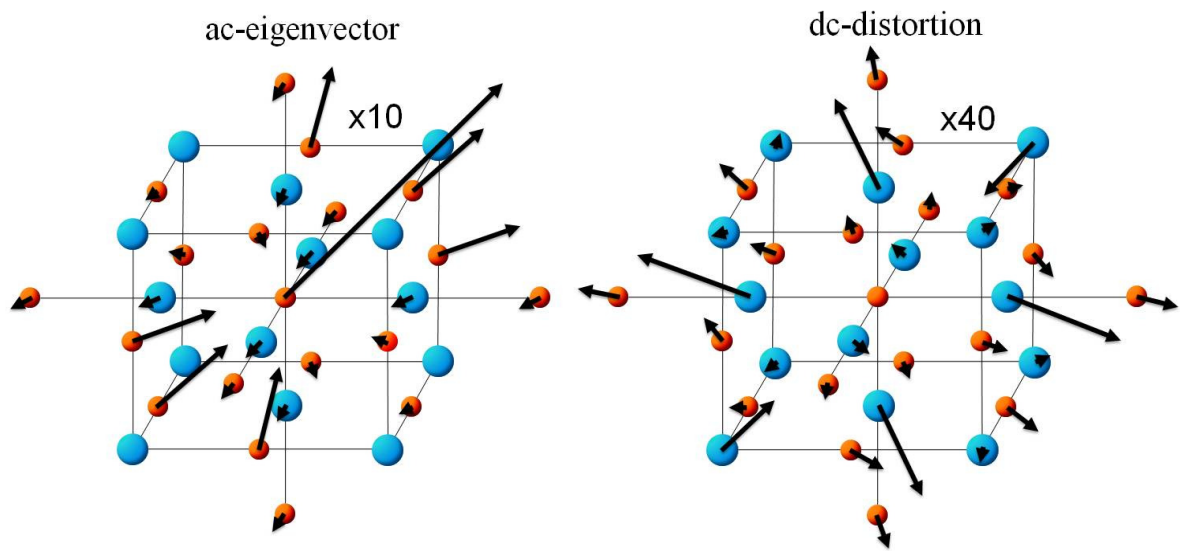

FIG. 1

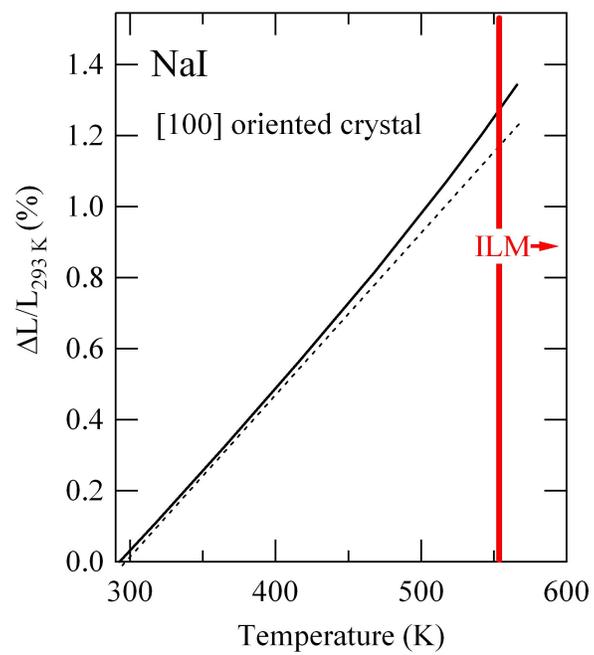

FIG. 2

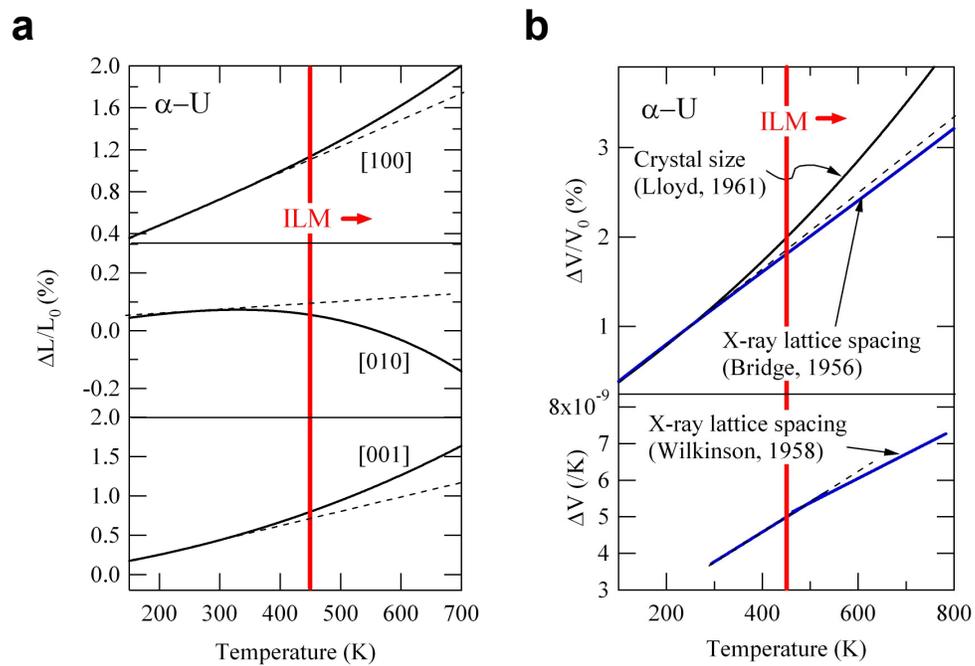

FIG. 3

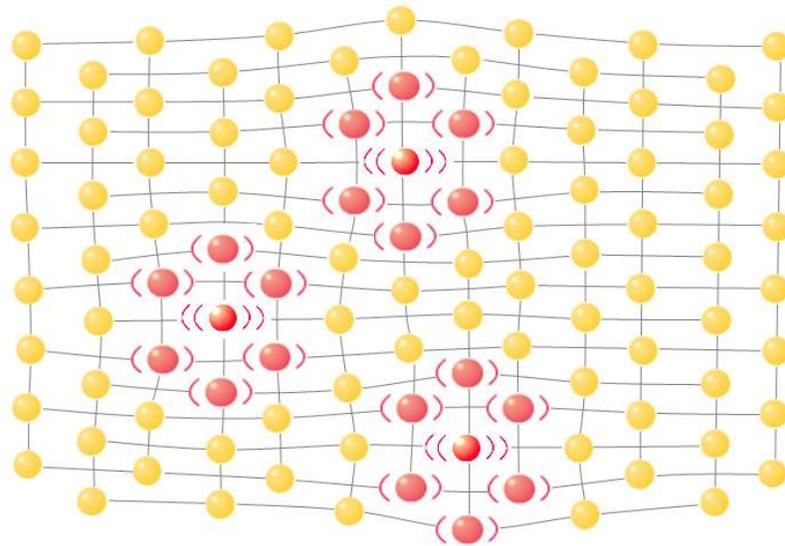

FIG. 4

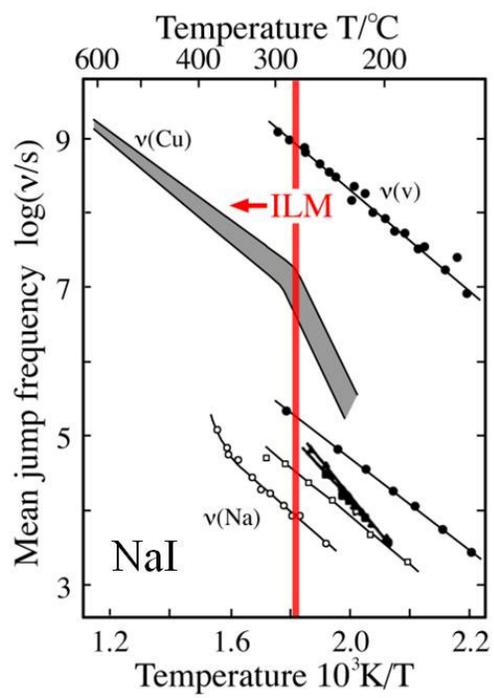

FIG. 5

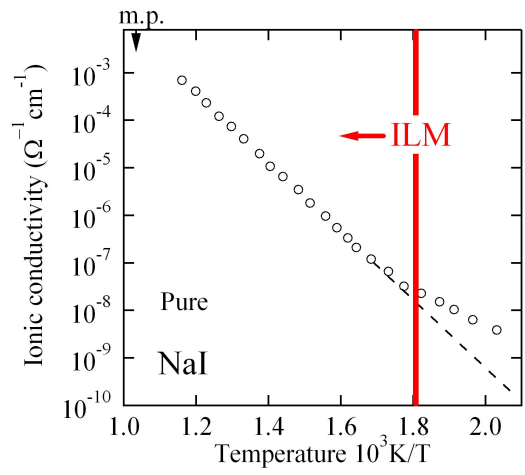 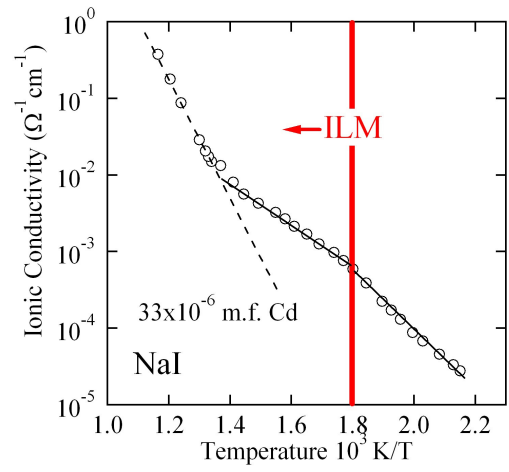

FIG. 6

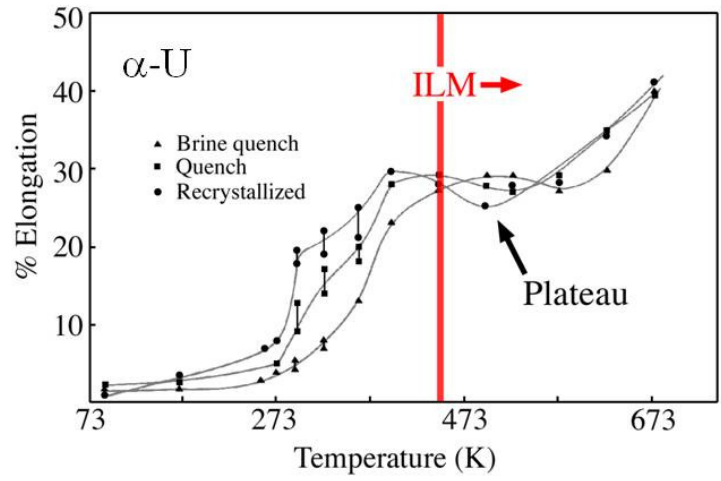

FIG. 7

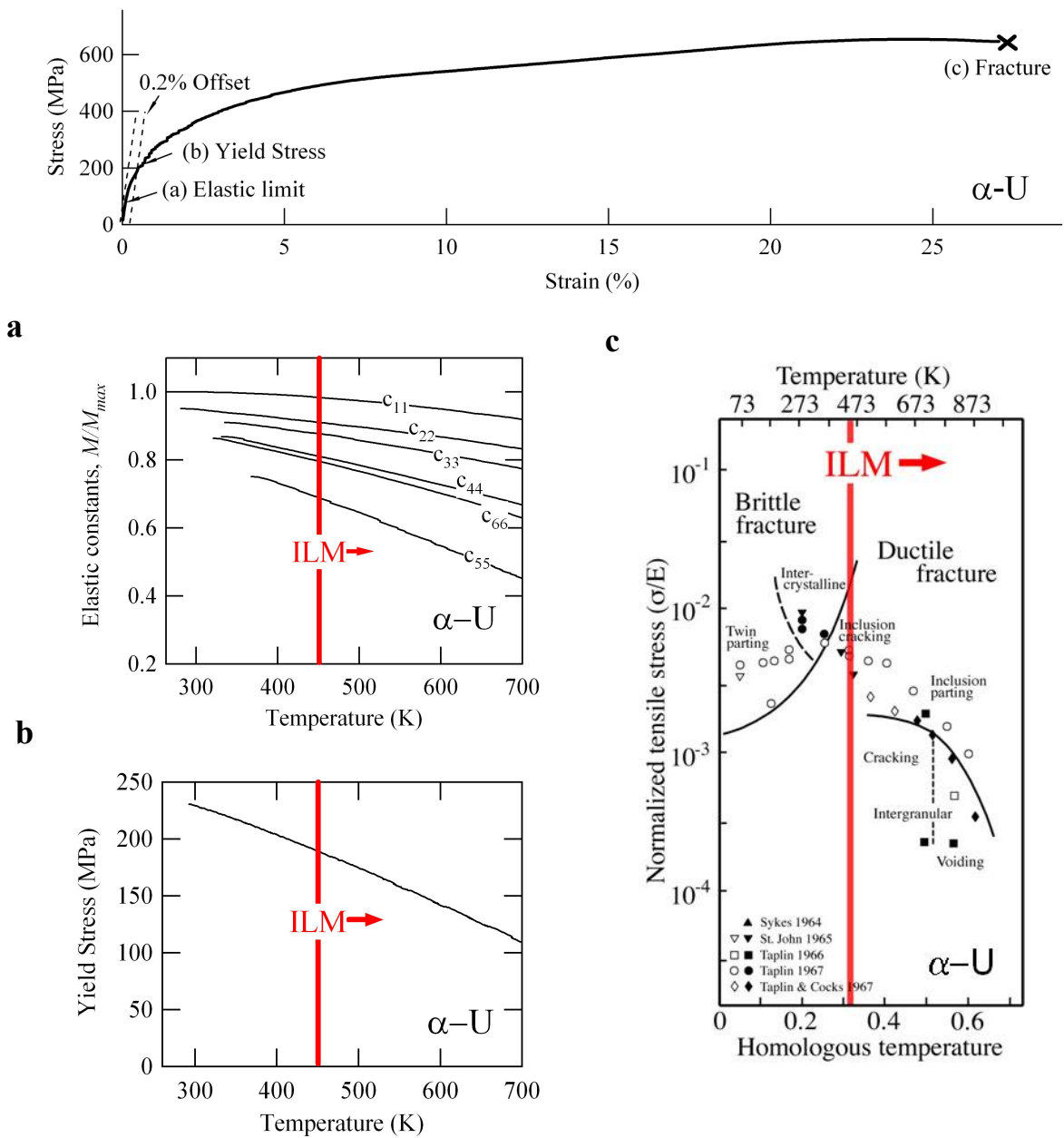

FIG. 8

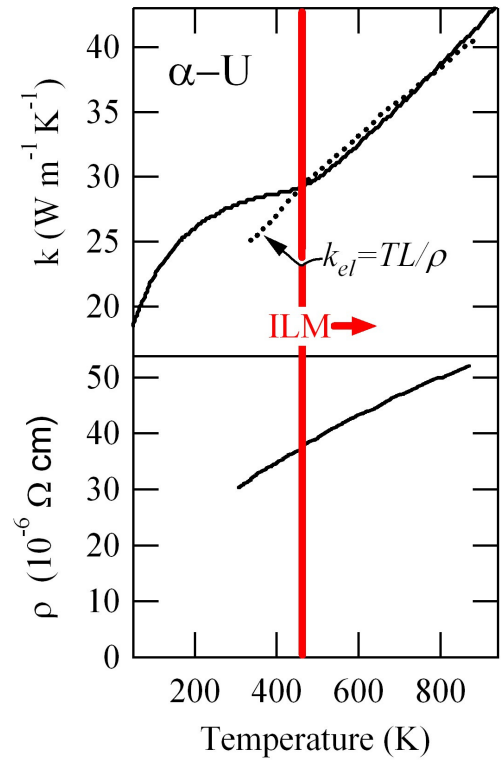

FIG. 9

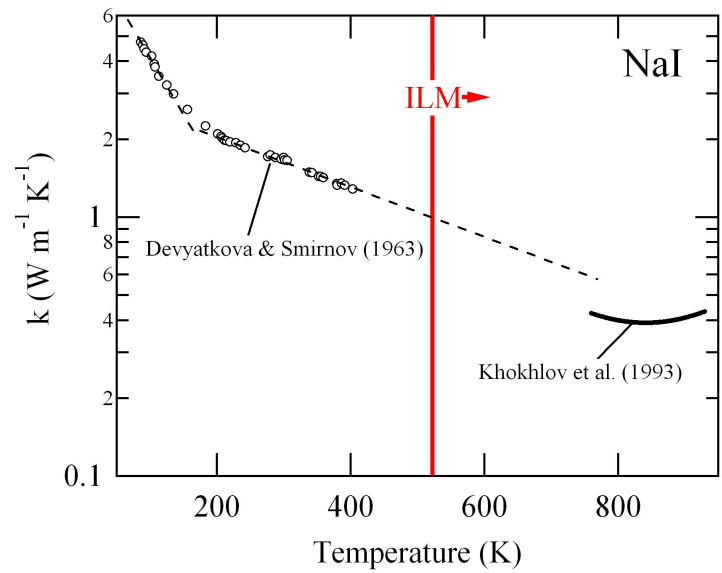

Fig. 10

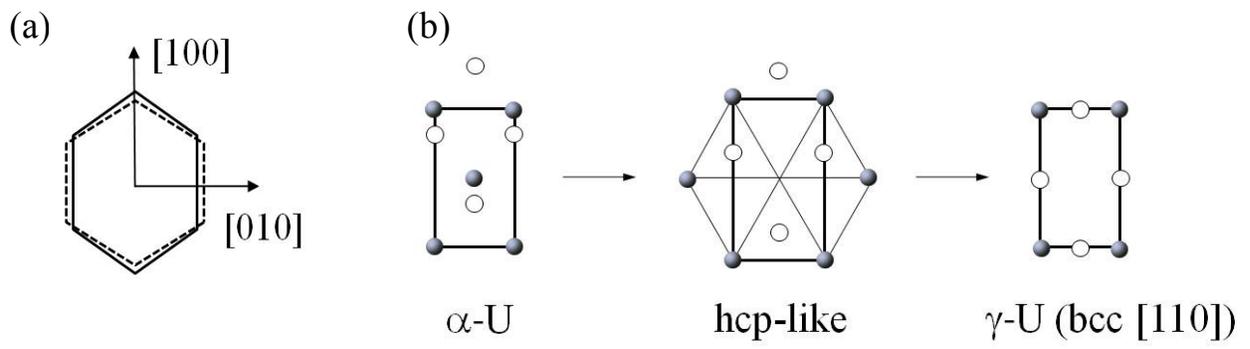

FIG. 11